\documentclass[a4paper,11pt,reqno]{amsart}
\usepackage{amsmath,amsthm,amssymb}

\numberwithin{equation}{section}

\theoremstyle{definition}
\newtheorem{df}{Definition}[section]
\newtheorem{prop}[df]{Proposition}

\newtheorem{rem}[df]{Remark}

\newtheorem{fact}[df]{Fact}

\title{Existence and  Orthogonality of Generalized Jack Polynomials and Its $q$-Deformation}
\author{Yusuke Ohkubo}
\address{Graduate School of Mathematics, Nagoya University, Nagoya 464-8602, Japan}
\email{m12010t@math.nagoya-u.ac.jp}

\begin{document}

\begin{abstract}
We investigate the existence and the orthogonality of the generalized Jack symmetric functions 
which play an important role in the AGT relations 
\cite{morozov2013finalizing}. 
We show their orthogonality by deforming them to the generalized Macdonald symmetric functions \cite{awata2011notes}. 
\end{abstract}

\maketitle

\section{Introduction}

In \cite{alday2010liouville} Aldey, Gaiotto and Tachikawa discovered surprising relations 
(called AGT relations) 
between 2D CFTs and 4D gauge theories. 
Since then, a number of studies on AGT relations have been carried out 
by mathematicians and physicists. 
One of them, 
a q-analogue of these relations is given by \cite{AwataYamada1, AwataYamada2}, 
in which connections between the deformed Virasoro/W algebra 
and 5D gauge theories are proposed. 
Moreover in \cite{awata2011notes}, 
it is conjectured with the help of the Ding-Iohara algebra 
that the matrix elements of the vertex operators 
with respect to the generalized Macdonald symmetric functions 
reproduce the Nekrasov factors of the 5D instanton partition function. 
These generalized Macdonald functions are the eigenfunctions of a certain operator 
in the Ding-Iohara algebra. 
At last, it is shown that the correlation functions of the Ding-Iohara algebra 
coincide with the 5D instanton partition functions in \cite{awata2012quantum}. 

On the other hand, 
the 4D $SU(2)$ AGT relation as Hubbard-Stratanovich (HS) duality is proved in \cite{morozov2013finalizing} 
with the help of the generalized Jack symmetric functions. 
(The formula of their selberg averages is still conjecture.)
In their previous paper \cite{beta=1}, 
the integrand of the Dotsenko-Fateev (DF) representation of the 4-point conformal block 
is expanded by the ordinary Jack symmetric polynomials 
and compared with the Nekrasov partition function. 
They show that, in the case $\beta=1$, 
every term of the DF-integral coincides with the Nekrasov formula 
parametrized by pairs of Young diagrams. 
However at $\beta \neq 1$ 
it is not working. 
At this, 
a generalization of the Jack symmetric polynomials is introduced in \cite{morozov2013finalizing} 
such that each term of DF-integral expanded by the generalized Jack polynomials 
corresponds to each term of Nekrasov partition function. 
However, 
the existence of the generalized Jack symmetric functions is unproven. 
Also, it is a little hard 
to give a mathematical, general proof 
of the orthogonality \cite[Section 2]{SU(3)GnJack} of these functions, 
because
they have degenerate eigenvalues. 

The generalized Macdonald and Jack symmetric functions are introduced independently 
and their relation has not been known. 
However, 
we find that in the limit $q \rightarrow 1$, 
the generalized Macdonald symmetric functions 
are reduced to the generalized Jack symmetric functions naturally. 
This follows the orthogonality and the existence 
of the generalized Jack symmetric functions. 
Since the eigenvalues of the generalized Macdonald symmetric functions are non-degenerate, 
we can prove their orthogonality 
without any other commutative operators, 
i.e., higher rank hamiltonians. 
Moreover 
we may bescribe 5D AGT relations as HS duality 
by using the generalized Macdonald symmetric functions of \cite{awata2011notes} 
in the same way of \cite{morozov2013finalizing}. 

This letter is organized as follows. 
In section 2, 
we give a short summary of the generalized Macdonald symmetric functions. 
In section 3, 
we calculate the limit $q \rightarrow 1$ of 
the operator whose eigenfunctions are the generalized Macdonald functions 
and prove 
the existence and the orthogonality of the generalized Jack symmetric functions. 
In section 4 we present several examples. 

\section{Generalized Macdonald symmetric function}

The generalized Macdonald symmetric functions are described 
by $N$ kinds of independent variables $\{ x_n^{(i)} \mid n \in \mathbb{N}, i = 1,\ldots , N \}$. 
Hence we use $N$ kinds of power sum symmetric functions $p^{(i)}_n= \sum_{k\geq 1} \left( x^{(i)}_k \right)^n$. 
Using this notation,  set 
\begin{align}
 &\eta^{(i)} (z) 
  \mathbin{:=} \exp \left( \sum_{n=1}^{\infty} \frac{1-t^{-n}}{n} z^{n} p^{(i)}_n \right) 
               \exp \left( -\sum_{n=1}^{\infty}(1-q^n) z^{-n} \frac{\partial }{\partial p^{(i)}_n} \right), \\
 &\xi^{(i)} (z) 
  \mathbin{:=} \exp \left(- \sum_{n=1}^{\infty} \frac{1-t^{-n}}{n}(t/q)^{\frac{n}{2}} z^{n} p^{(i)}_n \right) 
               \exp \left( \sum_{n=1}^{\infty} (1-q^n) (t/q)^{\frac{n}{2}} z^{-n} \frac{\partial }{\partial p^{(i)}_n} \right), \\
 &\varphi^{(i)}_+(z) 
  \mathbin{:=} \exp \left( - \sum_{n=1}^{\infty} (1-q^n) (1-t^n q^{-n}) (t/q)^{-\frac{n}{4}} z^{-n} \frac{\partial }{\partial p^{(i)}_n} \right), \\
 &\varphi^{(i)}_-(z) 
  \mathbin{:=} \exp \left( \sum_{n=1}^{\infty} \frac{1-t^{-n}}{n} (1-t^n q^{-n}) (t/q)^{-\frac{n}{4}} z^{n} p^{(i)}_n \right) , 
\end{align}
where $q$ and $t$ are independent parameters. 
Then these operators represent the Ding-Iohara algebra \cite{FHHSY}. 
Moreover for complex parameters $u_i$ $(i=1,2,\ldots , N)$ we define
\begin{equation}
X(z) \mathbin{:=} \sum_{i=1}^{N} u_i\, \widetilde{\Lambda}_i ,
\end{equation}
\begin{equation}
\widetilde{\Lambda}_i \mathbin{:=} \varphi^{(1)}_-((t/q)^{\frac{1}{4}} z)\,   \varphi^{(2)}_-((t/q)^{\frac{3}{4}} z) \, \cdots \, \varphi^{(i)}_-((t/q)^{\frac{2i-3}{4}} z)\, \eta^{(i)}((t/q)^{\frac{i-1}{2}}z). 
\end{equation}
The generalized Macdonald functions are the eigenfunctions of the operator
\begin{equation}
X_0 \mathbin{:=} \oint \frac{dz}{2 \pi \sqrt{-1} z} X(z).
\end{equation}

The generalized Macdonald symmetric functions are parametrized 
by $N$-tuples of partitions $\vec{\lambda}=(\lambda^{(1)},\ldots , \lambda^{(N)})$, 
where a partition $\lambda^{(i)}=(\lambda^{(i)}_1, \lambda^{(i)}_2, \ldots) $ 
is a sequence of non-negative integers such that 
$\lambda^{(i)}_1 \geq \lambda^{(i)}_2 \geq \cdots \geq 0$. 
For each partition $\lambda$ 
and an $N$-tuple of partitions $\vec{\lambda}=(\lambda^{(1)}, \ldots , \lambda^{(N)})$, 
we use the symbols : 
$|\lambda| \mathbin{:=} \sum_{k \geq 1} \lambda_k$, 
$\ell (\lambda) \mathbin{:=} \# \{k \mid \lambda_k \neq 0 \}$ 
and $|\vec{\lambda}| \mathbin{:=} \sum_{i=1}^N |\lambda^{(i)}| $. 
We write $\vec{\lambda} \geq^{\mathrm{L}} \vec{\mu}$ 
(resp. $\vec{\lambda} \geq^{\mathrm{R}} \vec{\mu}$) 
if and only if  $|\vec{\lambda}|=|\vec{\mu}|$ and
\begin{equation}
|\lambda^{(N)}|+ \cdots + |\lambda^{(j+1)}|+\sum_{k=1}^i \lambda^{(j)}_k 
\geq |\mu^{(N)}|+ \cdots + |\mu^{(j+1)}|+\sum_{k=1}^i \mu^{(j)}_k
\end{equation}
\begin{equation}
\left( \mathrm{resp. } \quad  |\lambda^{(1)}|+ \cdots + |\lambda^{(j-1)}|+\sum_{k=1}^i \lambda^{(j)}_k 
\geq |\mu^{(1)}|+ \cdots + |\mu^{(j-1)}|+\sum_{k=1}^i \mu^{(j)}_k \right)
\end{equation}
for all $i\geq 1$ and $1 \leq j \leq N$. 
Then "$\geq^{\mathrm{L}}$"  and "$\geq^{\mathrm{R}}$" 
are generalized dominace partial orderings on the $N$-tuples of partitions. 
By using these partial orderings 
we can triangulate the representation matrix of $X_0$. 
\begin{fact}[\cite{awata2011notes}]\label{fact:triangurate X_0}
Let $m_{\vec{\lambda}}$ be the product of monomial symmetric functions 
$\prod_{i=1}^{N} m_{\lambda^{(i)}}^{(i)}$ 
$( m_{\lambda^{(i)}}^{(i)}$ is an usual monomial symmetric function 
of variables $ \{x^{(i)}_n \mid n \})$ 
and 
$\mathbb{F}$ be $\mathbb{Q}(q^{\frac{1}{4}},t^\frac{1}{4})$. 
Then 
\begin{equation}
X_0 m_{\vec{\lambda}}=\sum_{\vec{\mu} \leq^{\mathrm{L}} \vec{\lambda}} c_{\vec{\lambda} \vec{\mu}} m_{\vec{\mu}}, \quad d_{\vec{\lambda} \vec{\mu}} \in \mathbb{F}(u_1, \ldots , u_N)
\end{equation}
and the eigenvalues are
\begin{equation}
c_{\vec{\lambda} \vec{\lambda}} 
= \sum_{i=1}^N u_i \, \epsilon_{\lambda^{(i)}} , \quad \epsilon_{\lambda}= 1+(t-1)\sum_{k=1}^{\ell(\lambda)}(q^{\lambda_k}-1)t^{-k}.
\end{equation}
\end{fact}
Thus we have the following existence theorem. 
\begin{fact}[\cite{awata2011notes}]
For each $N$-tuple of partitions $\vec{\lambda}$, 
there exists a unique symmetric function $P_{\vec{\lambda}}$ 
satisfying the following two conditions: 
\begin{align}
 &P_{\vec{\lambda}}=m_{\vec{\lambda}}+\sum_{\vec{\mu} <^{\mathrm{L}} \vec{\lambda}} d_{\vec{\lambda} \vec{\mu}}\, m_{\vec{\mu}} , \qquad d_{\vec{\lambda} \vec{\mu}} \in \mathbb{F}(u_1, \ldots , u_N); \\
 &X_0 P_{\vec{\lambda}}= e_{ \vec{\lambda}} P_{\vec{\lambda}} ,\qquad e_{ \vec{\lambda}} \in \mathbb{F}(u_1, \ldots , u_N). 
\end{align}
\end{fact}
In the case $N=1$, 
the symmetric functions $P_{\vec{\lambda}}$ are the usual Macdonald symmetric functions. 
Hence we call these symmetric functions $P_{\vec{\lambda}}$ 
the generalized Macdonald symmetric functions. 

The partial ordering "$\geq^{\mathrm{R}}$" 
gives 
the similar existence theorem of the dual symmetric functions $P_{\vec{\lambda}}^*$ 
for the adjoint operator $X_0^*$ of $X_0$
with respect to Macdonald's scalar product $\left\langle -,- \right\rangle_{q,t}$ 
defined by 
\begin{equation}
\left\langle p_{\vec{\lambda}}, p_{\vec{\mu}} \right\rangle_{q,t}
=\delta_{\vec{\lambda}, \vec{\mu}} \prod_{i=1}^{N} z_{\lambda^{(i)}}
 \prod_{k=1}^{\ell (\lambda^{(i)})}\frac{1-q^{\lambda_k^{(i)}}}{1-t^{\lambda_k^{(i)}}} , \qquad
z_{\lambda^{(i)}} \mathbin{:=} \prod_{k \geq 1}k^{m_k} \, m_k !, 
\end{equation}
where $m_k$ is the number of entries in $\lambda^{(i)}$ equal to $k$ and 
$p_{\vec{\lambda}} \mathbin{:=} \prod_{i=1}^{N}p_{\lambda^{(i)}}^{(i)} \mathbin{:=} \prod_{i=1}^{N} \prod _{k \geq 1} p_{\lambda^{(i)}_k}^{(i)}$. 
When $q$ and $t$ are generic, 
the eigenvalues $e_{\vec{\lambda}}$ of the generalized Macdonald symmetric functions are non-degenerate, 
that is 
\begin{equation}
\vec{\lambda} \neq \vec{\mu} \quad \Rightarrow \quad e_{\vec{\lambda}} \neq e_{\vec{\mu}}. 
\end{equation}
Therefore we have the following orthogonality of them. 
\begin{fact}[\cite{awata2011notes}]\label{fact:orthogonality of GnMac}
If $\vec{\lambda} \neq \vec{\mu}$ then
\begin{equation}
\left\langle P^*_{\vec{\lambda}}, P_{\vec{\mu}} \right\rangle_{q,t} =0.
\end{equation}
\end{fact}

\section{Limit to $\beta$ deformation}

In this section, 
we set $u_i=q^{u_i'}$ 
$( i=1, \ldots , N )$, 
$t=q^{\beta}$, 
$q=e^{\hbar}$ 
and take the limit $\hbar\rightarrow 0$ with $\beta$ fixed 
in order to consider the specialization to $\beta$-deformation. 
Since $(1-t^{-n})(1-t^n q^{-n})/n=\mathcal{O}(\hbar^2)$ and 
$(1-t^{-n})/n$, $(1-q^{n})=\mathcal{O}(\hbar)$, 
\begin{align}
\oint \frac{dz}{2\pi \sqrt{-1} z}\widetilde{\Lambda}_i(z) 
 =&1+ \sum_{n=1}^{\infty} \left( -\frac{(1-t^{-n})(1-q^n)}{n} p_{n}^{(i)} \frac{\partial }{\partial p_{n}^{(i)}} \right) \\
  &+ \sum_{k=1}^{i-1} \sum_{n=1}^{\infty} \left( -\frac{(1-t^{-n})(1-t^n q^{-n})(1-q^n)}{n} (t/q)^{-\frac{(i-k-1)n}{2}} p_{n}^{(k)} \frac{\partial }{\partial p_{n}^{(i)}} \right) \nonumber \\
  &+ \frac{1}{2}\sum_{n,\, m} \left( -\frac{(1-t^{-n})(1-t^{-m})(1-q^{n+m})}{n\, m } p_{n}^{(i)} p_{m}^{(i)} \frac{\partial }{\partial p_{n+m}^{(i)}} \right) \nonumber \\
  &+ \frac{1}{2}\sum_{n,\, m} \left( \frac{(1-t^{-n-m})(1-q^{n})(1-q^{m})}{ (n+m)} p_{n+m}^{(i)} \frac{\partial }{\partial p_{n}^{(i)}} \frac{\partial }{\partial p_{m}^{(i)}} \right) + \mathcal{O}(\hbar^4).  \nonumber
\end{align}
Hence the $\hbar$ expansion is
\begin{align}
\oint \frac{dz}{2\pi \sqrt{-1} z}\widetilde{\Lambda}_i(z) 
=& 1+ \hbar^2 \left\{ \beta \sum_{n=1}^{\infty} n \, p^{(i)}_n \frac{\partial }{\partial p^{(i)}_n} \right\} 
 + \hbar^3 \left\{ \beta (1-\beta) \sum_{k=1}^{i-1} \sum_{i=1}^{\infty} n^2 p^{(k)}_n \frac{\partial }{\partial p^{(i)}_n} \right.  \\
 & + \frac{\beta ^2}{2}\sum_{n, m} (n+m) p^{(i)}_n p^{(i)}_m \frac{\partial }{\partial p^{(i)}_{n+m}} 
 + \frac{\beta }{2}\sum_{n, m} n\, m\, p^{(i)}_{n+m} \frac{\partial^2 }{\partial p^{(i)}_n \partial p^{(i)}_m }  \nonumber \\
 &+ \left. \frac{\beta(1-\beta)}{2} \sum_{n=1}^{\infty} n^2 p^{(i)}_n \frac{\partial }{\partial p^{(i)}_n}  \right\} + \mathcal{O}(\hbar^4).  \nonumber
\end{align}
Thus we get
\begin{align}
u_i \oint \frac{dz}{2\pi \sqrt{-1} z}\widetilde{\Lambda}_i(z) 
=& 1+ u_i' \, \hbar + \hbar^2 \left\{ \beta \sum_{n=1}^{\infty} n \, p^{(i)}_n \frac{\partial }{\partial p^{(i)}_n} +\frac{1}{2}u_i'^2 \right\} \\
 & + \hbar^3 \left\{ \beta\,  \mathcal{H}_{\beta}^{(i)} + \beta \sum_{k=1}^{i-1} \mathcal{H}_{\beta}^{(i,k)} + \frac{u_i'^3}{6}  \right\} + \mathcal{O}(\hbar^4),  \nonumber
\end{align}
where
\begin{equation}
\mathcal{H}_{\beta}^{(i)}
\mathbin{:=} \frac{1}{2}\sum_{n,m} \left( \beta (n+m) p^{(i)}_n p^{(i)}_m \frac{\partial }{\partial p^{(i)}_{n+m}}+n\, m\, p^{(i)}_{n+m} \frac{\partial^2 }{\partial p^{(i)}_n \partial p^{(i)}_m } \right) 
 +\sum_{n=1}^{\infty} \left( u_i' + \frac{1-\beta}{2} n \right) n p^{(i)}_n \frac{\partial }{\partial p^{(i)}_n}, 
\end{equation}
\begin{equation}
\mathcal{H}_{\beta}^{(i,k)} \mathbin{:=} (1-\beta)\sum_{n=1}^{\infty} n^2 p^{(k)}_n \frac{\partial }{\partial p^{(i)}_n}.
\end{equation}
For $k=0,1,2,\ldots$ ,
we define operators $H_k$ by 
\begin{equation}
X_0 \mathbin{=:} \sum_{k=0}^{\infty} \hbar^k H_k .
\end{equation}
With respect to  $H_0$, $H_1$ and $H_2$, 
all homogeneous symmetric functions $m_{\vec{\mu}}$ in $P_{\vec{\lambda}}$ 
belong to the same eigenspace. 
Even without the operators at these levels, 
the eigenfunctions of $X_0$ do not change. 
In addition, we have
\begin{equation}
\lim_{\hbar \rightarrow 0} \left( \frac{X_0-(H_0+\hbar H_1+ \hbar^2 H_2)}{(t-1)(q-1)^2}\right) =\mathcal{H}_{\beta} +\frac{1}{6 \beta}\sum_{i=1}^{N}u_i'^3 , 
\end{equation}
\begin{equation}
\mathcal{H}_{\beta} \mathbin{:=} \sum_{i=1}^{N} \mathcal{H}_{\beta}^{(i)} + \sum_{i>j}\mathcal{H}_{\beta}^{(i,j)}. 
\end{equation}
Consequently the limit $q \rightarrow 1$ of the generalized Macdonald functions 
are eigenfunctions of the differential operator $\mathcal{H}_{\beta}$. 
As a matter of fact, 
$\mathcal{H}_{\beta}$ 
plus the momentum $(\beta-1)\sum_{i=1}^{N} \sum_{n =1}^{\infty} n p^{(i)}_n \frac{\partial }{\partial p^{(i)}_n}$
corresponds to the differential operator of \cite{morozov2013finalizing, SU(3)GnJack}, 
the eigenfunctions of which are called generalized Jack symmetric functions\footnote{
To be adjusted to the notation of \cite{morozov2013finalizing, SU(3)GnJack}, 
we need to transform the subscripts:
$u_i' \rightarrow u_{N-i+1}'$, $p^{(i)} \rightarrow p^{(N-i+1)}$.
}. 

As the Fact \ref{fact:triangurate X_0} 
we can triangulate $\mathcal{H}_{\beta}$ similarly. 
Moreover 
if $\vec{\lambda} \geq^{\mathrm{L}} \vec{\mu}$ 
and $\beta$ is generic, 
then $e'_{\vec{\lambda}} \neq e'_{\vec{\mu}}$. 
($e'_{\vec{\lambda}}$, $e'_{\vec{\mu}}$ are eigenvalues of $\mathcal{H}_{\beta}$.)
Therefore we get the existence theorem of the generalized Jack symmetric functions. 
\begin{prop} 
There exists a unique symmetric function $J_{\vec{\lambda}}$ 
satisfying the following two conditions: 
\begin{align}
 &J_{\vec{\lambda}}=m_{\vec{\lambda}}+\sum_{\vec{\mu} <^{\mathrm{L}} \vec{\lambda}} d'_{\vec{\lambda} \vec{\mu}}\, m_{\vec{\mu}} , \qquad d'_{\vec{\lambda} \vec{\mu}} \in \mathbb{Q}(\beta, u'_1,\ldots ,u'_N); \\
 &\mathcal{H}_{\beta} J_{\vec{\lambda}}= e'_{ \vec{\lambda}} J_{\vec{\lambda}} ,\qquad e'_{ \vec{\lambda}} \in \mathbb{Q}(\beta , u'_1,\ldots ,u'_N). 
\end{align}
\end{prop}
From the above argument and the uniqueness in this proposition
we get the following important result. 
\begin{prop}\label{prop:limit of GnMac}
The limit of the generalized Macdonald symmetric functions $P_{\vec{\lambda}}$ 
to $\beta$-deformation coincide with the generalized Jack symmetric functions $J_{\vec{\lambda}}$. 
That is 
\begin{equation}
P_{\vec{\lambda}} \underset{\substack{\hbar \rightarrow 0,\\ u_i=q^{u_i'},\, t=q^{\beta},\, q=e^{\hbar}}}{\longrightarrow} J_{\vec{\lambda}}.
\end{equation}
\end{prop}
\begin{rem}
For the dual functions $P_{\vec{\lambda}}^*$ and $J_{\vec{\lambda}}^*$, 
the similar proposition holds. 
\end{rem}
By Fact \ref{fact:orthogonality of GnMac}, 
Proposition \ref{prop:limit of GnMac} and 
the fact that the scalar product $\left\langle -,- \right\rangle_{q,t}$
reduces to the scalar product $\left\langle -,- \right\rangle_{\beta}$
which is defined by 
\begin{equation}
\left\langle p_{\vec{\lambda}}, p_{\vec{\mu}} \right\rangle_{\beta} 
=\delta_{\vec{\lambda}, \vec{\mu}} \prod_{i=1}^{N} z_{\lambda^{(i)}} \beta^{-\ell (\lambda^{(i)})}, 
\end{equation} 
we obtain the orthogonality of the generalized Jack symmetric functions. 
\begin{prop}
If $\vec{\lambda} \neq \vec{\mu}$, 
then 
\begin{equation}
\left\langle J_{\vec{\lambda}}^* , J_{\vec{\mu}}  \right\rangle_{\beta} = 0. 
\end{equation}
\end{prop}

\section{Example}

We give examples of Proposition \ref{prop:limit of GnMac} in the case $N=2$.
The generalized Macdonald symmetric functions of level 1 and 2
have the forms:
\begin{equation}
\left(
\begin{array}{l}
P_{(0),(1)}  \\
P_{(1),(0)} 
\end{array}
\right)
=
M_{q,t}^1 
\left(
\begin{array}{l}
m_{(0),(1)}  \\
m_{(1),(0)} 
\end{array}
\right), \qquad
M_{q,t}^1 
\mathbin{:=}
\left(
\begin{array}{cc}
 1 & (t/q)^{\frac{1}{2}} \frac{(t-q)u_2}{t(u_1-u_2)} \\
 0 & 1
\end{array}
\right), 
\end{equation}
\begin{equation}
\left(
\begin{array}{l}
P_{(0),(2)}  \\
P_{(0),(1,1)} \\
P_{(1),(1)}  \\
P_{(2),(0)} \\
P_{(1,1),(0)}  
\end{array}
\right)
=
M_{q,t}^2
\left(
\begin{array}{l}
m_{(0),(2)}  \\
m_{(0),(1,1)} \\
m_{(1),(1)}  \\
m_{(2),(0)} \\
m_{(1,1),(0)} 
\end{array}
\right), 
\qquad M_{q,t}^2 \mathbin{:=}
\end{equation}
{\footnotesize 
\begin{equation*}
 \left(
\begin{array}{ccccc}
1 & \frac{(1+q)(t-1)}{q t-1} &  \frac{(t/q)^{-\frac{1}{2}} (1+q)(q-t)(t-1)u_2}{(1-q t)(u_1-q u_2)} & \frac{(q-t)((1-q^2)t u_1 - q(t^2 -q(1+q)t+q)u_2 ) u_2}{q t (q t -1)(u_1 - u_2 )(u_1- q u_2)} & \frac{(1+q)(q-t)(t-1)((q-1)t u_1 +q(q-t)u_2)u_2}{qt (qt-1)(u_1-u_2)(u_1-q u_2)}  \\
0 & 1 & (t/q)^{\frac{1}{2}} \frac{(t-q)u_2}{t(t u_1 -u_2)} & \frac{(q-t)u_2}{q(t u_1 -u_2)} & \frac{(q-t)(q u_2 - t((t-1)u_1+u_2))u_2}{q t (u_1 -u_2)(tu_1 -u_2)} \\
0 & 0 & 1 & (t/q)^{\frac{1}{2}} \frac{(t-q)u_2}{t(qu_1-u_2)} & (t/q)^{\frac{1}{2}} \frac{(q-t)((1+q+(q-1)t)u_1-2t u_2)}{t(q u_1- u_2)(-u_1 +t u_2)} \\
0 & 0 & 0 & 1 & \frac{(1+q)(t-1)}{q t-1} \\
0 & 0 & 0 & 0 & 1
\end{array}
\right)
\end{equation*}
}

\noindent Also the generalized Jack symmetric functions have the forms: 
\begin{equation}
\left(
\begin{array}{l}
J_{(0),(1)}  \\
J_{(1),(0)} 
\end{array}
\right)
=
M_{\beta}^1 
\left(
\begin{array}{l}
m_{(0),(1)}  \\
m_{(1),(0)} 
\end{array}
\right), \qquad 
M_{\beta}^1 
\mathbin{:=}
\left(
\begin{array}{cc}
 1 & \frac{1-\beta}{-u'_1+u'_2} \\
 0 & 1
\end{array}
\right), 
\end{equation}
\begin{equation}
\left(
\begin{array}{l}
J_{(0),(2)}  \\
J_{(0),(1,1)} \\
J_{(1),(1)}  \\
J_{(2),(0)} \\
J_{(1,1),(0)}  
\end{array}
\right)
=
M_{\beta}^2
\left(
\begin{array}{l}
m_{(0),(2)}  \\
m_{(0),(1,1)} \\
m_{(1),(1)}  \\
m_{(2),(0)} \\
m_{(1,1),(0)} 
\end{array}
\right),
\end{equation}
\begin{equation}
M_{\beta}^2 \mathbin{:=}
 \left(
\begin{array}{ccccc}
 1 &\frac{2\beta}{1+\beta} & \frac{2\beta(1-\beta)}{(1+\beta)(1-u'_1+u'_2)} & \frac{(1-\beta)(2+\beta-\beta^2-2u'_1+2u'_2)}{(1+\beta)(u'_1-u'_2)(-1+u'_1-u'_2)} & \frac{2\beta(2-3\beta+\beta^2)}{(1+\beta)(u'_1-u'_2)(-1+u'_1-u'_2)}  \\
 0 &1 &\frac{1-\beta}{-\beta-u'_1+u'_2} &\frac{1-\beta}{\beta+u'_1-u'_2} &\frac{-1+3\beta-2\beta^2}{(u'_1-u'_2)(-\beta-u'_1+u'_2)}  \\
 0 &0 &1 & \frac{1-\beta}{-1-u'_1+u'_2} & \frac{2(1-\beta)(-1+\beta-u'_1+u'_2)}{(-1-u'_1+u'_2)(\beta-u'_1+u'_2)} \\
 0 &0 &0 &1 &\frac{2\beta}{1+\beta}  \\
 0 &0 &0 &0 &1  
\end{array}
\right). 
\end{equation}
If we take the limit $q \rightarrow 1$ of $M_{q,t}^i$, 
then $M_{\beta}^i$ appears. 

\section*{Acknowledgements}

The author would like to show his deepest gratitude to his supervisor H. Awata 
for his enormous help 
and also thanks H. Kanno and A. Morozov very much for their valuable comments. 


\end{document}